# Three-Dimensional Continuous Multi-Walled Carbon Nanotubes Network-Toughened Diamond Composite

Jiawei Zhang[1,2], Keliang Qiu[3], Tengfei Xu[4], Xi Shen[1], Junkai Li[5], Fengjiao Li[1], Richeng Yu[1], Huiyang Gou[5], Duanwei He[6], Liping Wang[7], Zhongzhou Wang[3], Guodong Li[8], Yusheng Zhao[2], Ke Chen[3,8,*], Fang Hong[1,*], Ruifeng Zhang[4,*], Xiaohui Yu[1,*]

1. Beijing National Laboratory for Condensed Matter Physics, Institute of Physics, Chinese Academy of Sciences, Beijing 100190, P. R. China.
2. Eastern Institute for Advanced Study, Eastern Institute of Technology; Ningbo 315201, P. R. China.
3. School of Chemistry, Beihang University; Beijing 100191, P. R. China.
4. School of Materials Science and Engineering, Beihang University; Beijing 100191, P. R. China.
5. Center for High Pressure Science and Technology Advanced Research; Beijing 100193, P. R. China.
6. Institute of Atomic and Molecular Physics, Sichuan University; Chengdu 610065, P. R. China.
7. Academy for Advanced Interdisciplinary Studies, and Department of Physics, Southern University of Science and Technology; Shenzhen 518055, P. R. China.
8. State Key Lab of Tropic Ocean Engineering Materials and Materials Evaluation, Hainan University, Haikou 570228, P. R. China.
9. These authors contributed equally: Jiawei Zhang, Keliang Qiu, Tengfei Xu, Xi Shen,

**Email**: yuxh@iphy.ac.cn; chenke0119@buaa.edu.cn; hongfang@iphy.ac.cn; zrf@buaa.edu.cn

**Enhancing the fracture toughness of diamond while preserving its hardness is a significant challenge. Traditional toughening strategies have primarily focused on modulating the internal microstructural units of diamonds, including adjustments to stacking sequences, faults, nanotwinning, and the incorporation of amorphous phases, collectively referred to as intrinsic toughening. Here, we introduce an extrinsic toughening strategy to develop an unparalleled tough diamond composite with complex and abundant $sp^2$-$sp^3$ bonding interfaces, by incorporating highly dispersed multi-walled carbon nanotubes (MWCNTs) into the gaps of diamond grains to create a three-dimensional (3D) continuous MWCTNs network-toughen heterogeneous structure. The resultant composite exhibits a hardness of approximately 91.6 GPa and a fracture toughness of roughly 36.4 MPa·m$^{1/2}$, which is six times higher than that of synthetic diamond and even surpasses that of tungsten alloys, surpassing the benefits achievable through intrinsic toughening alone. The remarkable toughening behavior can be attributed to the formation of numerous mixed $sp^2$-$sp^3$ bonding interactions at the 3D continuous network MWCNTs/diamond interfaces, which facilitate efficient energy dissipation. Our 3D continuous network heterogeneous structure design provides an effective approach for enhancing the fracture toughness of superhard materials, offering a new paradigm for the advanced composite ceramics.**

Diamond with unique tetrahedral arrangement of $sp^3$-bonded carbon atoms, is a quintessential engineering material (e.g., cutting tools) that boasts extremely high hardness (60–120 GPa), excellent thermal conductivity, and optical transparency[1,2]. However, its low fracture toughness (3.4–5.0 MPa·m$^{1/2}$)[3,4] renders it inherently prone to brittle fracture, thereby limiting its broader industrial applications[5]. Enhancing diamond's fracture toughness typically comes at the expense of its hardness[6], making the simultaneous improvement of both properties a significant challenge due to their trade-off relationship. In recent decades, several studies intrinsic toughening strategies have been developed to enhance the fracture toughness of diamond without sacrificing hardness through nanostructure engineering, including nanotwinned diamond architectures[6], graphite–diamond hybrids[7], diamond–graphene composites[8], and amorphous diamond phases[9]. For instance, constructing nanotwin boundaries (NTBs) in diamonds has been shown to be superior to large-angle grain boundaries in enhancing their mechanical responses, particularly in fracture toughness[6,10-12]. The formation of high-density nanotwins—crystalline regions related by symmetry—can significantly toughen diamond[6]. Recently, a hierarchically structured architecture in diamond, featuring coherently interfaced polytypes, interwoven nanotwins, and interlocked nanograins, has achieved a fracture toughness of up to 26.6 MPa·m$^{1/2}$, surpassing what can be achieved through nanotwinning alone[5]. In other hard composites such as ceramics and glass[13], intrinsic toughening strategies have also been reported to improve fracture toughness, e.g. bio (nacre, enamel)-inspired structure[14], temper[15], and phase engineering[16]. However, despite these advances, the challenge of significantly enhancing the fracture toughness of diamond without compromising its high hardness by intrinsic toughening strategies remains unresolved.

Carbon nanotubes (CNTs) -- cylindrical derivatives of graphene ($sp^2$-bonded honeycomb lattice) derivatives -- possess excellent mechanical properties and a high aspect ratio, making them ideal candidates for extrinsic reinforcements for toughening diamonds over the past two decades[17,18]. However, the extrinsic toughening mechanisms of CNTs within diamond matrices remain poorly understood and highly debated, with most experimental results falling significantly short of the theoretical predictions when CNTs—either single- or multi-walled—are introduced into various matrices[19]. A key challenge lies in establishing efficient interfacial interactions between the CNTs and the matrix, as the fracture toughness is highly dependent on the strength of the interface and the load transfer across it[20,21]. The unique chemical bonds inherent in CNTs often result in lower sliding resistance, akin to the super-lubricity of graphite, which undermines their toughening potential[22]. Therefore, optimizing interfacial bonding interactions is crucial for unlocking the full potential of CNTs in diamond composites, particularly for achieving dramatic improvements in toughness, which requires an atomistic understanding of the interface-dominated mechanical behaviour and deep insights into the underlying electronic mechanisms[23]. In addition to optimizing interfacial bonding, maintaining the high dispersibility and uniformity of CNTs within the diamond matrix is equally pivotal. Creating a uniform hetero-phase structural arrangement is essential for promoting positive mechanical responses at multiscale levels, yet this remains longstanding and critical challenge[5,24,25]. Addressing these two challenges of interfacial bonding and CNT dispersion motivate the development of advanced toughening strategies to overcome the brittleness of diamonds, thereby facilitating their broader commercialization across various applications.

To endow diamonds with unprecedented fracture toughness, one avenue worth exploring involves constructing a three-dimensional (3D) continuous network-toughen architecture in a diamond matrix material. For this purpose, we proposed an extrinsic toughening strategy. This strategy introduces highly-dispersed multi-walled carbon nanotubes (MWCNTs) into the gaps of diamonds to construct a 3D continuous MWCNTs network in the polycrystal diamond matrix (3D-MWCNTs-diamond) during high-temperature and high-pressure (HTHP) condition, forming a uniform and compact heterogeneous structure, where the MWCNTs are efficiently bonded with diamond at the MWCNTs/diamond interfaces through multiple bonding interactions such as $sp^2$-$sp^3$ covalent bonds, Van der Waals' force and physical contacts. Single-edge notched beam (SENB) tests of 3D-MWCNTs-diamond achieve an impressive toughness as high as 36.4 MPa·m$^{1/2}$, six times that of synthetic diamond and even surpasses that of tungsten alloys. Further in situ fracture testing in TEM, combined with DFT simulation, demonstrates two modes of crack propagation. When entering regions of relatively weak $sp^2$ bonding 3D-MWCNTs/diamond interfaces, the crack propagation path is sinuous and wavy, accompanied by MWCNTs deformation (pull out) and the interface migration. Within diamond phase, the crack propagates rapidly along {111} planes in a straight manner.

**Design, preparation and characterization**

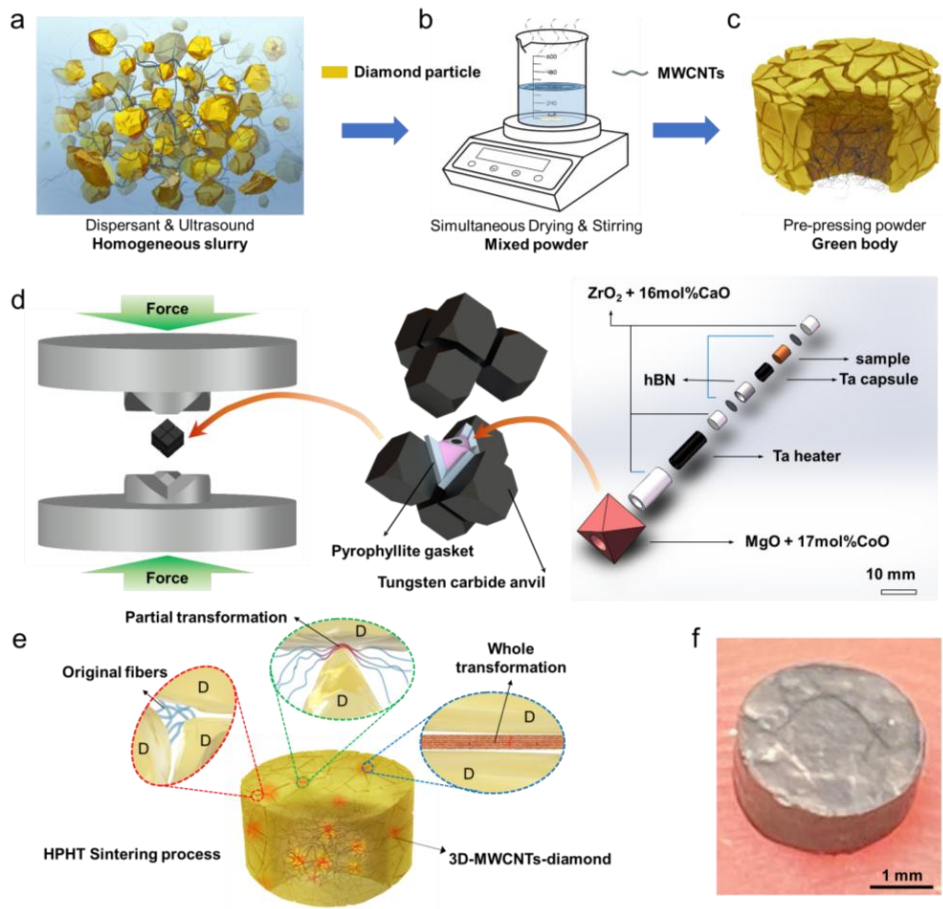

**Fig. 1 | Preparation process and structural features of 3D-MWCNTs-diamond composite**. **a-c,** Schematic for a homogeneous slurry of MWCNTs and diamond (a), the preparation of highly dispersed

MWCNTs-diamond composite powder (b), and the production of a crude green body assembled from the composite powder (c). **d**, Multilevel, exploded schematics of a two-stage multi-anvil apparatus based on a Walker-type press, showing the preparation of the composite. **e**, Schematic for 3D-MWCNTs-diamond composite with multiple structural features. Three enlarged oval insets show three existing states of MWCNTs in the composite: Red, original fiber; Green, partial transformation into diamond; Blue, whole transformation into diamond. Earthy yellow grains represent diamond grains, black skeleton represents MWCNTs, and D indicates diamond. **f**, An optical photograph of the composite bulk (size: a diameter of 3 mm and thickness of 2 mm).

The three-dimensional (3D) continuous MWCTNs network-toughen diamond (named as 3D-MWCNTs-diamond) heterogeneous structure can be intelligently designed and constructed, as schematically shown in Fig. 1a-e. Firstly, a certain amount of MWCNTs fibers are uniformly dispersed in diamond grains to prepare a homogeneous slurry by ultrasonic dispersion (Fig. 1a). Next, the water in the slurry is slowly evaporated to produce a mixed power by simultaneous heating and stirring process, realizing the highly dispersed MWCNTs fibers into the gaps between diamond grains and on the surfaces of diamond grains for avoiding the agglomeration of MWCNTs in diamond matrix (Fig. 1b and Supplementary Fig. 1). Subsequently, a green body can be obtained by pre-pressing the prepared mixed MWCNTs-diamond power to assemble a cylindrical bulk as a precursor (Fig. 1c). Ultimately, a 3D-MWCNTs-diamond composite can be constructed through transferring the cylindrical precursor into the two-stage multi-anvil apparatus under HTHP conditions (Fig. 1d). Owing to the inherent ultra-high strength of diamonds, there often exist in uneven stress distributions between compressed diamond grains in HTHP condition[26,27]. Notably, when the preparation temperature is high enough (≥2000 °C), the diamond strength can be partly reduced and the stress deviation will be weakened to some extent[26]. Therefore, when the preparation condition is rationally optimized, primarily including appropriate sintering temperature and suitable pressure, a structurally perfected 3D-MWCNTs-diamond composite can be schematically achieved, as shown in Fig. 1e, where a low-pressure cavity supported by diamond grains protects MWCNTs from being transformed into diamond, a high stress zone can induce the contact point/surface extrusion between diamond grains which is beneficial to form strong covalent bonding interactions between MWCNTs and diamond or between diamonds (because of partial or full transformation of MWCNTs into diamond).

To achieve the structurally perfected 3D-MWCNTs-diamond composite, we studied the optimal sintering temperature (800~2200 °C) under a pressure of 15 GPa. Through observing the fractured surfaces of these samples sintered at various temperatures, we found that, at lower sintering temperatures (800~1800 °C), the densification of the sintered samples was primarily through the fracture of diamond grains (Supplementary Fig. 2a-d); as the sintering process progressed, the contact area between diamond particles increased, and stress relaxation occurred, causing the composite's densification to approach its limit; however, a significant number of pores persisted, which helps to protect the MWCNTs from excessive pressure and compression. As the sintering temperature surpassed 2000 °C, the diamond grains started to undergo significant plastic deformation, which reduced the pores in the sintered samples and enabled further densification and diamond-diamond bonding (Supplementary Fig. 2e,f). The X-Ray Diffraction (XRD) spectra further reveal that the graphite peak of the composite vanishes when the sintering temperature exceeds 1800 °C (Supplementary Fig. 3). This disappearance is attributed to the decrease in diamond strength at higher temperatures, which accelerates the plastic deformation of diamond, resulting in an increase in the actual pressure within diamond grain gaps and the transformation of some compressed MWCNTs into diamond[26]. The absence of graphite peaks in the XRD spectrum of the 2000 °C composite, despite the observation of MWCNTs by STEM, may be due to the presence of only a small amount of graphite phase. A typical cylindrical 3D-MWCNTs-diamond composite bulk can be successfully prepared at 2000 °C and 15 GPa, as shown in Fig. 1f.

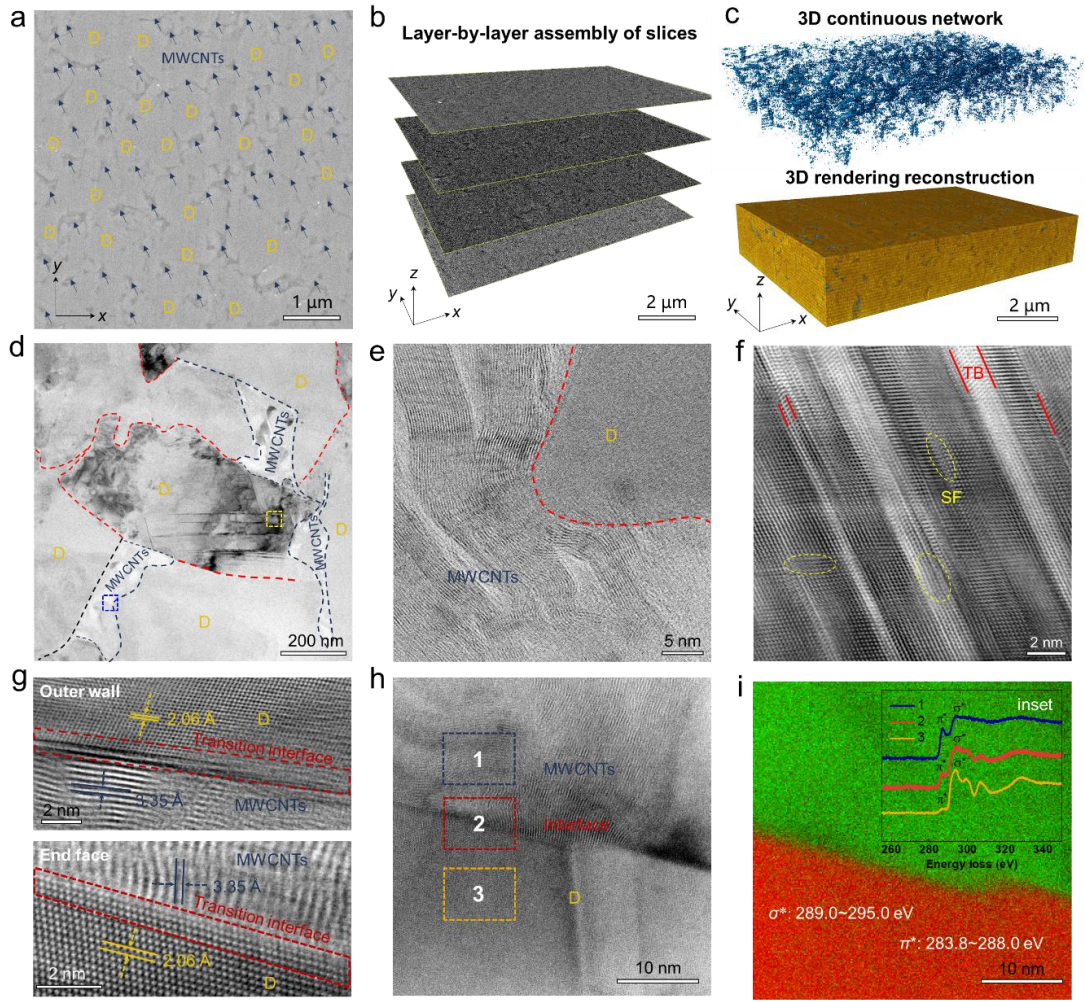

**Fig. 2 | Multi-scale structural characterizations of 3D-MWCNTs-diamond composite prepared at 2000 °C and 15 GPa.** **a**, The SEM image of a typical composite slice, obtained by focused ion beam (FIB) technology. Dark blue arrows indicate the gaps around diamond particles. D represents diamond. **b**, The layer-by-layer assembled image of the four slices on the $x \times y$ plane (size: 10 μm × 6 μm × 20 nm). Dark spots indicate MWCNTs. **c**, 3D rendering reconstruction image of MWCTNs dispersed in diamond (top) and 3D-MWCNTs-diamond composite (bottom), by analyzing typical SEM images of the sixty composite slices. **d**, Low-magnified ABF-STEM image of the 3D-MWCNTs-diamond, showcasing that MWCNTs fill the gaps between diamond grains. **e**, Enlarged STEM image of hybrid MWCNTs-diamond interface, corresponding to the blue box in (d), showing that MWCNTs extruded

by diamond grains bond well with adjacent diamonds or phase transform into diamonds. **f**, High resolution STEM image of diamond defect structures from MWCNTs phase transitions, corresponding to the white box in (**d**), showing that MWCNTs fully transform into diamond nanotwin bands. SF denotes stacking fault marked in yellow and TB represents twin boundaries. **g**, Atomic structure of typical heterogeneous MWCNTs/diamond interfaces in the composite characterized by ABF-STEM, viewed along [110]$_{Diamond}$ direction. The bonding interaction between MWCNTs and diamond occurs at the end face (bottom) or outer wall (top) of MWCNTs. **h,i**, Low-magnified HADDF-STEM image of the MWCNTs/diamond interfacial structure (h), corresponding to its EELS mapping of π* (green) and σ* (red) characteristic peaks (**i**). inset (i): EELS spectra of different regions in **h** (dark blue box: MWCNTs, dark red box: the interface, orange: diamond). The peaks at 285 and 292 eV can be attributed to transitions of a C$_{1s}$ electron to π* and σ* states, corresponding to *sp²* and *sp³* bonding, respectively. The coexistence of red and green at the interface indicates the presence of a transition interface connected by mixed *sp²* and *sp³* bonding.

We performed scanning electron microscopy (SEM), combined with 3D rendering reconstruction technology, and scanning transmission electron microscopy (STEM) analysis, to systemically characterize the tailored 3D-MWCNTs-diamond structure at multi-scale levels, as shown in Fig. 2 and Supplementary Figs. 4 and 5. The SEM images of the typical composite slices show a uniform defect network (dark regions) around diamond particles on 2D plane (Fig. 2a and Supplementary Fig. 4), where the MWCNTs fibers as second phase are fully restricted to these defect regions. The typical assembled image of the continuous four slices further verifies that the heterogeneous structure is also uniform (Fig. 2b). Based on reconstructing the continuous sixty slices to form 3D rendering models (10 μm × 6 μm × 1.2 μm) by Avizo software, we reconfirm that the 3D continuous MWCNTs network structure can be perfectly formed (Figure 2c, top), and is uniformly embedded into the diamond matrix for creating an integral unit (Figure 2c, bottom). Simultaneously, the volume fraction of the MWCNTs is estimated to be about 0.35% by analyzing the 3D reconstruction model. A low-magnification annular bright-field (ABF) STEM image demonstrates that a tangled, embedded 3D network skeleton of MWCNTs is perfectly constructed in the gaps of diamond grains (Fig. 2d). Simultaneously, we also observe that another portion of the same MWCNTs fiber situated in the low-pressure cavity retains its original structure (Supplementary Fig. 5). This continuity of the MWCNTs fiber across both regions enables diamond grain bridging, thus forming the expected 3D continuous MWCNTs skeleton structure. Besides, the enlarged STEM image confirms that a portion of a MWCNTs fiber located in a high stress zone bonds well to the adjacent diamond grains (Fig. 2e). Meanwhile, the high-resolution STEM image verifies that partial MWCNTs can transform into diamond nanotwin bands in the high stress region after HPHT sintering (Fig. 2f)[28].

In addition, the representative ABF-STEM images reveal that the 3D-MWCNTs-diamond sample exhibits a highly coherent arrangement at the MWCNTs-diamond interface. The coherent bonding interactions between MWCNTs and diamond occurs at either the end face (Fig. 2g, bottom; and Supplementary Fig. 5c) or the outer wall (Fig 2g, top; and Supplementary Fig. 2d) of the MWCNTs. The bonding is primarily achieved by random *sp³* (the end face) or *sp²* (the outer wall) hybrid covalent bonds formed under HPHT, similar to the transition interface from graphite to diamond[29]. Meanwhile, a typical electron energy loss spectroscopy (EELS) mapping at the transition interface verifies the presence of the coherent interface with mixed $sp^2$–$sp^3$ bonding (Fig. 2h,i). This transition interface, connected through mixed $sp^2$–$sp^3$ bonding, can be formed not only by the direct bonding of MWCNTs to diamond but also by the phase boundary of MWCNTs fibers undergoing a partial phase transition, the middle section under high-stress compression is easily transformed into diamond.

**Mechanical performance**

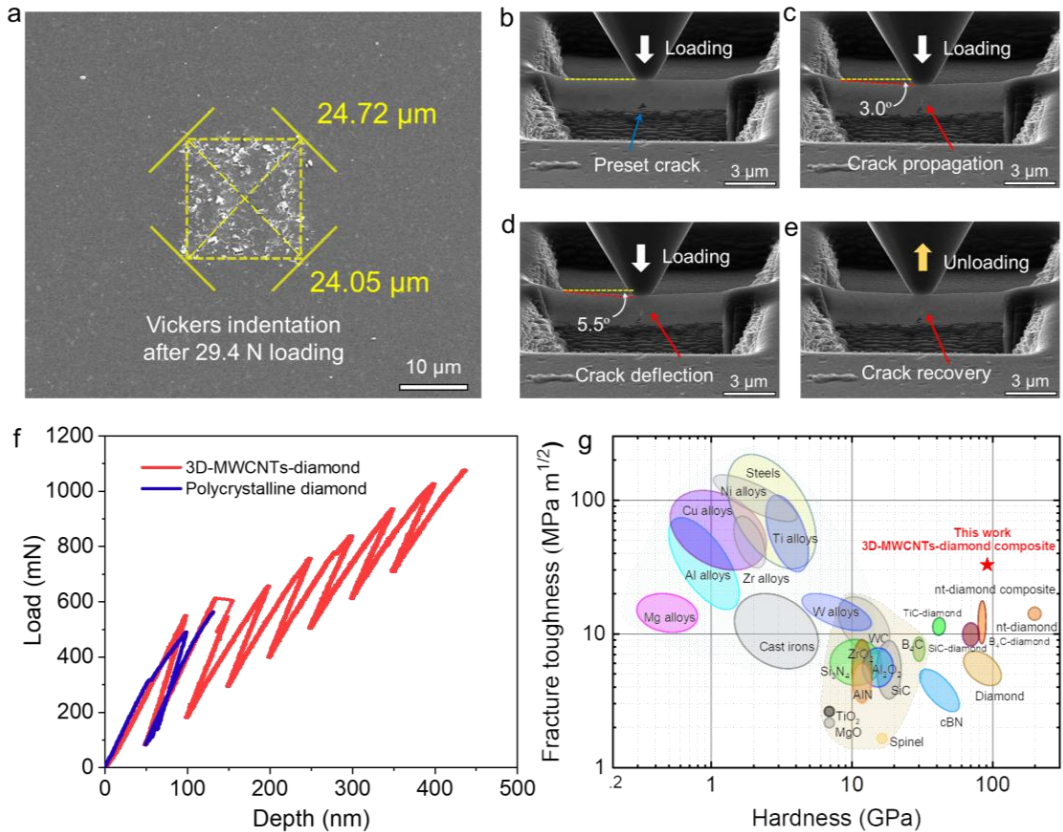

**Fig. 3 | Mechanical performance of 3D-MWCNTs-diamond composite. a**, Representative post-indentation SEM image of the composite surface, subjected to a 29.4 N load by a Vickers indentation test, indicating no visible indentation crack. **b-e**, Sequential snapshots of the composite micro-beam during the SENB test, with panel d corresponding to the time point when the maximum diamond indenter displacement is reached, just before the beam fractures completely. **f**, Typical force-displacement curve of micro-beams of 3D-MWCNTs-diamond composite and polycrystalline diamond, obtained by the SENB test. **g**, Ashby plot of the hardness against fracture toughness for the composite compared with typical engineering metals, ceramics, superhard materials, and other superhard composites.

We initially evaluate the fracture toughness on the free surface of the 3D-MWCNTs-diamond composite prepared at 15 GPa and 2000 °C, using an indentation fracture method[26]. Interestingly, after a large load (29.4 N) indented on the free surface of the composite, there are no cracks around the post-indentation (Fig. 3a), suggesting its high fracture toughness. We further perform the single-edge notched beam (SENB) tests on the composite sample to accurately estimate its fracture toughness (Fig. 3b-f, Supplementary Fig. 6 and Movie 1). We can observe that, as the top diamond indenter is pressed down, a crack initiates at the tip of the precut notch, first propagating to the upper left (Fig. 3c) and then turning nearly 45° and extending to the upper right (Fig. 3d). This large angle crack deflection consumes additional energy[30], suggesting the exceptional fracture toughness of 3D-MWCNTs-diamond composite. After retracting the probe (Fig. 3e), the crack heals to a considerable extent. The SENB test results demonstrate that the average fracture toughness of the composite micro-beam is $31.9 \pm 4.6$ MPa·m$^{1/2}$, maximally reaching up to 36.4 MPa·m$^{1/2}$ (as listed in Supplementary Table 1), which is about 6 times greater than that of single crystal diamond and 1.4 times higher than that of the toughest hierarchically structured diamond composite (26.6 MPa·m$^{1/2}$)[5]. Meanwhile, we confirm that the Vickers hardness of the composite is $91.6 \pm 3.1$ GPa under a 29.4 N load, comparable to that of single-crystal diamond[1,6]. Besides, we also measured the fracture toughness of pure polycrystalline diamond (Supplementary Fig. 7). The Vickers indentation toughness values ($12.4 \pm 2.5$ MPa·m$^{1/2}$) are lower than the SENB ones ($16.5 \pm 2.3$ MPa·m$^{1/2}$, by about 25%), consistent in previously reported difference[5]. Therefore, the relative difference in toughness between pure polycrystalline diamond and 3D-MWCNTs-diamond composite is analogous in these two methods. The estimated indentation fracture toughness of 3D-MWCNTs-diamond composite is approximately $24.8 \pm 2.5$ MPa·m$^{1/2}$. In comparison with the toughness *vs* hardness of some typical engineering metals, ceramics, superhard materials (e.g., diamond, nt-diamond), and other superhard composites (e.g., B$_4$C-diamond, TiC-diamond, nt-diamond composite)[5,6,35-40], we highlight the unique combination of fracture toughness and high hardness exhibited by the 3D-MWCNTs-diamond composite (Fig. 3f).

Before obtaining the optimal combination of fracture toughness and hardness of the composite, the maximal toughening can be achieved by controlling the sintering temperature from 1400 °C to 2200 °C, to form the special 3D continuous MWCNTs network structure under 15 GPa, as shown in Supplementary Fig. 7. We find that, as the sintering temperature gradually increases, the hardness of the composite also enhances step by step from $50.2 \pm 3.4$ GPa for 1400 °C to $91.6 \pm 3.1$ GPa for 2000 °C to $99.4 \pm 3.6$ GPa for 2200 °C, but the fracture toughness of the composite first increases from $8.0 \pm 3.4$ MPa·m$^{1/2}$ for 1400 °C to $31.9 \pm 4.6$ MPa·m$^{1/2}$ for 2000 °C, and then suddenly decreases to $12.4 \pm 2.3$ MPa·m$^{1/2}$ for 2200 °C (Supplementary Fig. 7a,b). Obviously, when a lower sintering temperature of 1400 °C is provided, good interface bonding is not sintered, resulting in a low hardness and low fracture toughness. However, when the temperature is further increased to 2200 °C, obvious cracks appear at the four corners of the Vickers indentation (Supplementary Fig. 7c), indicating its low fracture toughness. The toughening of 3D-MWCNTs-diamond weakens as the sintering temperature increases more than 2000 °C, because more MWCNTs transform into diamond[26], and the sample even gradually becomes transparent (inset in Supplementary Fig. 7c). In contrast, the hardness of 3D-MWCNTs-diamond almost linearly increases with the rising temperature as the sample densifies and diamond bonding strengthens. Finally, the optimum condition (at 15 GPa and 2000 °C) in preparation of 3D-MWCNTs-diamond for high hard and high fracture toughness was obtained from the above experiments. The aforementioned changes in hardness and toughness primarily result from the evolution of the composite's microstructure (Supplementary Fig. 2), phase composition (Supplementary Fig. 3), and interfacial bonding strength under HPHT conditions (discussed in the Supplementary Materials).

## Toughening mechanism

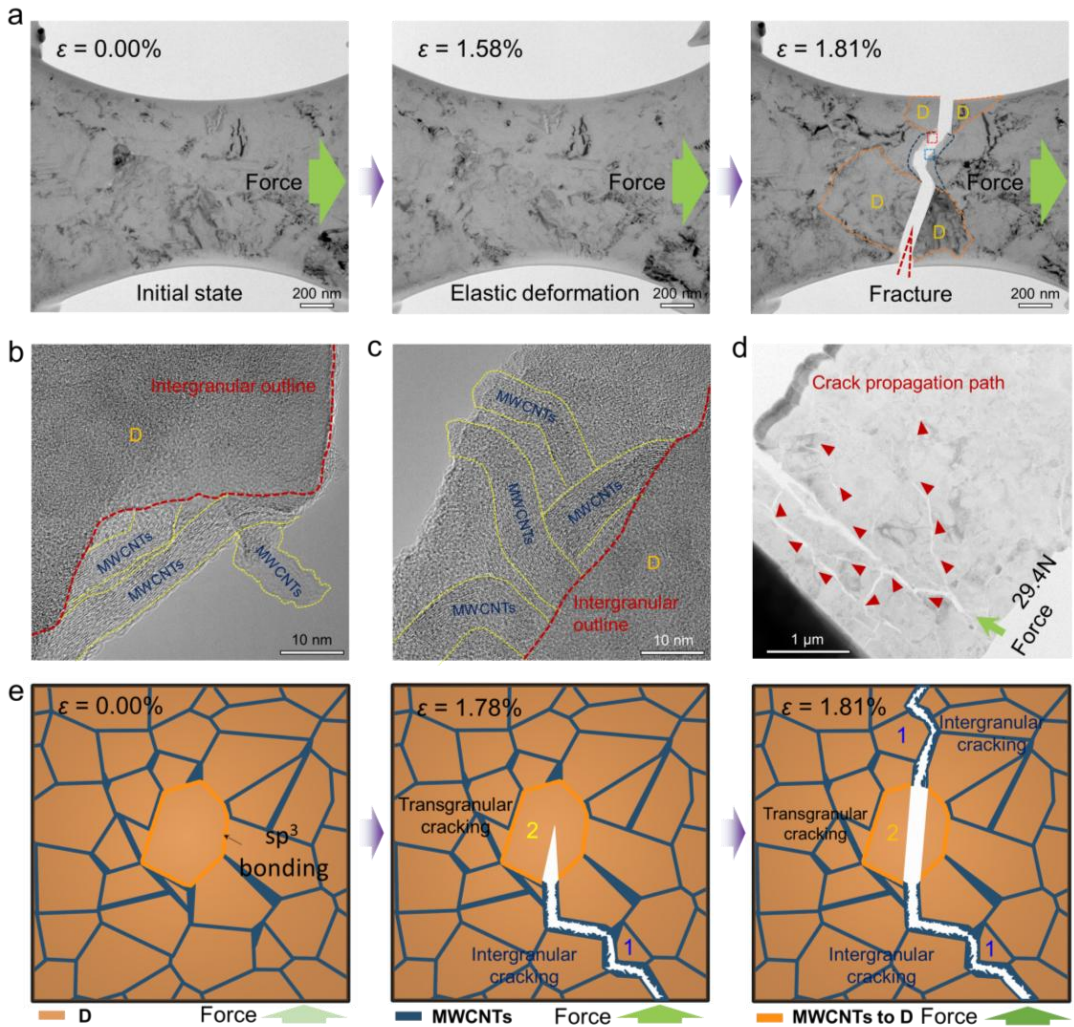

**Fig. 4 | Toughening mechanism of 3D-MWCNTs-diamond composite revealed by *in situ* TEM test. a**, Bright-field images of a composite slice at different tensile strains ($\varepsilon = 0$, 1.58%, and 1.81%). The orange dashed lines show a transgranular cracking region of the diamond grain. The dark blue dashed lines indicate the intergranular cracking regions. **b,c**, Enlarged TEM images, corresponding to the red and blue boxes in a, showing the pull-out (**b**) and bridging (**c**) of MWCNTs, respectively. Inset in b: MWCNTs pull out. Inset in c: MWCNTs bridged diamond interface. The dark red dashed lines represent the intergranular crack outline of the composite. **d**, Low-magnified ABF-STEM image of the composite microcracks under Vickers indentation, revealing a radiating multi-branched fracture path. Red arrows represent microcrack propagation paths. **e**, Sketch maps of this composite at three stages ($\varepsilon = 0$, 1.78%, and 1.81%) show the microcrack propagation paths of the composite more clearly during the tensile process. The weak bonding interfaces that separate the grains are denoted by the dark blue regions, and the strong covalent interfaces that bond the grains are represented by the orange regions. The

intergranular cracking region (1) that separated the grains is denoted by the frayed dark blue line and the transgranular cracking region (2) of the grain is denoted by the orange line.

To elucidate the toughening mechanism at the nanoscales, an *in-situ* tensile fracture test was first conducted in the transmission electron microscopy (TEM) on a specially designed 3D-MWCNTs-diamond composite slice, as shown in Fig. 4a and Supplementary Figs. 8-10. During the tensile fracture test, initially a stress concentration region occurred near the narrowest point (Fig. 4a, $\varepsilon = 0\%$), acting as a crack initiation source. With increasing strain, there was no crack initiation from the narrowest point ($\varepsilon = 1.58\%$). When the strain increased to about 1.75% (Supplementary Movie 2), a crack initiated near the narrowest point, instantaneously propagated until fracture suddenly at the strain of 1.81%. The whole crack propagation path in the region where MWCNTs are incorporated into the composite, exhibits frequent changes and deflections. These enlarged TEM images, corresponding to the different color boxes in Fig. 4a (Supplementary Fig. 8a, $\varepsilon = 1.81\%$) show that the cracks have sinuous and jagged propagating paths within the composite phase (1 region, Supplementary Fig. 8b), while it propagates along {111} plane in a straight manner within the diamond phase (2 region, consistent to the brittle fracture mode of the single diamond, Supplementary Fig. 8c). The multiple tensile experiments consistently demonstrated the special fracture mode in 3D-MWCNTs-diamond composite (Supplementary Figs. 9 and 10). This deflected behaviour of the composite may be attributed to the non-uniform internal stress distribution between the MWCNTs and the diamond matrix[26,27]. Because the relatively weak bonding interaction at the MWCNTs/diamond interface easily leads to crack deflections, contributing to the toughening effect[5]. Usually, the composite can sustain higher loads due to the load transfer effect arising from the mismatch in elastic moduli between the MWCNTs and the diamond matrix. Furthermore, higher-magnified TEM images demonstrated the MWCNTs were pulled out (Fig. 4b) and the diamond grains were bridged by MWCNTs (Fig. 4c) at the fracture regions. Obviously, the presence of MWCNTs (Fig. 2e) and diamond defect structures like stacking faults and diamond nanotwin bundles (Fig. 2f)[28], leads to sinuous and wavy boundaries with tapered fracture edges on each fractured surface. Therefore, the fracture edges of the composite exhibit a conical shape, which is a characteristic of the pull-out fracture mode. This fracture mode is conducive to the energy dissipation and toughening of a material[5], as it allows for a gradual release of stress and prevents sudden, catastrophic failure. Besides, typical STEM images show that the overall crack pattern exhibits radiating multi-branched fracture path on the composite surface under indentation (Fig. 4d and Supplementary Fig. 11), effectively dispersing the external load and minimizing the stress concentration. Actually, the large energy dissipation associated with large crack deflection, effectively reducing the local stress and strain for promoting the toughening of the composite. Based on the above experimental phenomenon, we propose that, as the force is initially applied, a microcrack easily initiates at the relatively weak interface of MWCNTs/diamond by $sp^2$ bonding; as the strain increases to 1.78%, a crack propagates and deflects along the diamond due to the strong interfacial hindrance, simultaneously, accompanied with the deformation (pull out) of MWCNTs, leading to an intergranular cracking (1 region); subsequently, when the crack extends to a diamond region, the crack propagating often traverses the entire diamond grain by a transgranular cracking mode (2 region), facilitating the dispersion and transfer of the load to the diamond grains interconnected by the strong $sp^3$ bonding of MWCNTs; with the increase of strain up to 1.81%, the crack continues to propagate along the interface of

MWCNTs/diamond until fully fracture by the intergranular cracking mode. Consequently, the dispersed and weakened crack propagation energy is consumed by the mixed cracking modes.

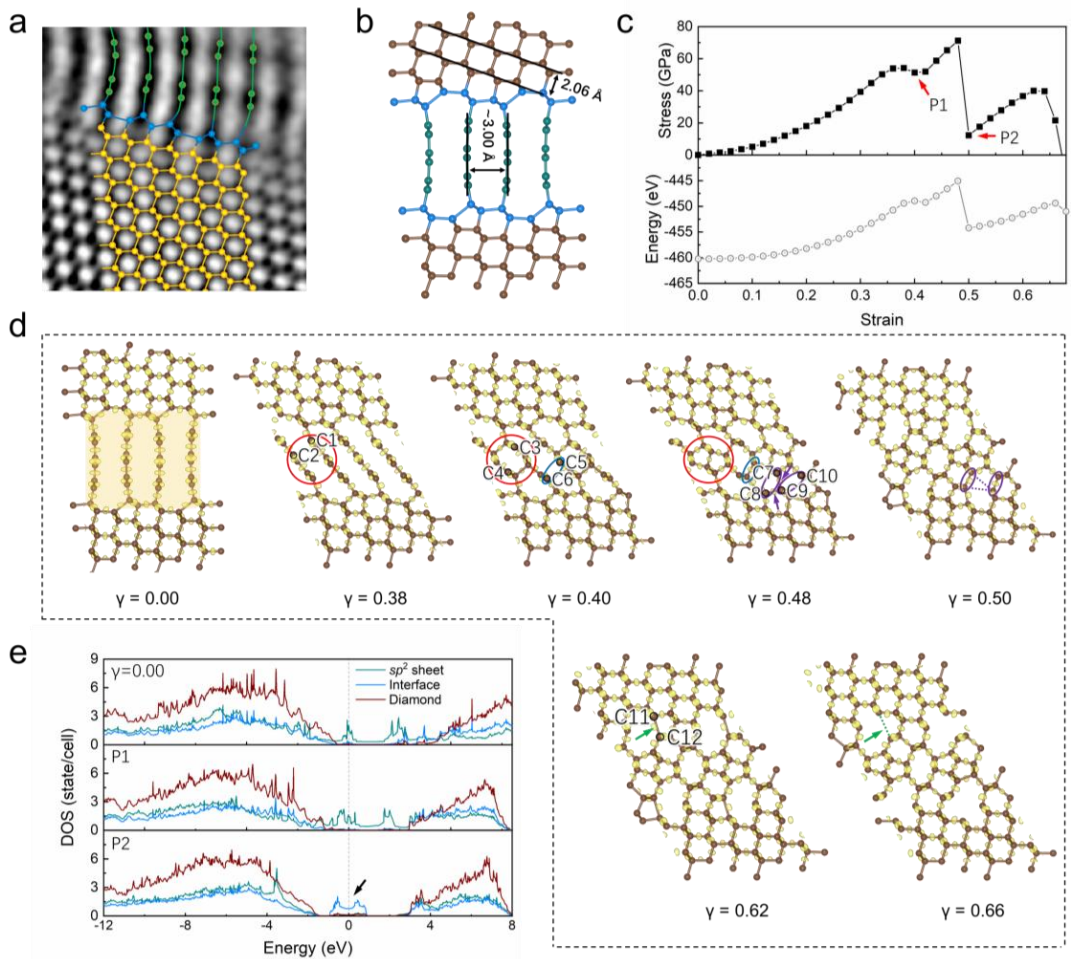

**Fig. 5 | DFT simulation results for simplified model of MWCNTs/diamond interface. a**, A ABF-STEM image of the coherent MWCNTs/diamond interface, matched with its ball-and-stick structural model. **b**, A structural model, in which the interface, $sp^2$ sheet and diamond-structured atoms are coloured in blue, green and brown, respectively. **c**, A calculated stress-strain of the structural model, corresponding to its energy-strain relationship under shear strain parallel to the interface. **d**, Structural snapshots of the structural model, corresponding to its electron localization function (ELF) maps at different key strains. **e**, The density of state (DOS) for the structural model at equilibrium conditions and the phase-transitioned polymorphs. The Fermi level is set to zero.

To elucidate the toughening mechanism at the atomic scale, we constructed a simplified model based on its heterogeneous structural bonding feature (Fig. 5a). This model comprises several layers of the diamond structure units and $sp^2$ sheets, which are reasonably reduced from $sp^2$ bonded MWCNT structures, allowing to investigate its plastic response under the shear strain parallel to the interface (Fig. 5b). Both the calculated stress-strain and energy-strain relationships (Fig. 5c) exhibit a sawtooth pattern, indicating multiple response stages, involving several bond breaks and changes in chemical bonding configurations[41-43]. Structural snapshots and the corresponding electron localization function (ELF) maps at key strains (equilibrium, several peaks and valleys, and failure strains) reveal the following sequence pf events. At a strain of 0.38 (the first peak value), the $sp^2$ sheets and the diamond portion deflect significantly. Subsequently at a strain of 0.40, the C1-C2 atoms located on the two columns of $sp^2$ sheets (highlighted in red circular regions) form bonds, resulting in the formation of a carbon polymorph P1. The stability of P1 is verified using ab initio molecular dynamics (AIMD) simulations (Supplementary Fig. 12). As the strain increases further to 0.48, the stress gradually rises to the second peak point, where the C3 and C4 atoms, along with the C5 and C6 atoms (marked in blue ellipses) located on the other two columns of $sp^2$ sheets are also bonded, forming a denser $sp^3$ lattice. At the strain of 0.50, the C7-C9 bond in the near-interfacial region breaks, causing a sharp drop in stress. This event leads to the reconfiguration of the surrounding atoms (highlighted in purple) and the formation of a new carbon polymorph P2, characterized by C7-C8 and C9-C10 bonding. Notably, P2 remains stable and continues to produce the next stage of plastic response until the fracture of the C11-C12 bond (marked in green arrows) at the strain of 0.66, leading to the final failure (Supplementary Fig. 12). The density of state (DOS) analysis indicates that, unlike the Gradia structure with $sp^2$–$sp^3$ mixed interface atoms[7], the metallic character of this composite primarily originates from the $sp^2$ sheets, while the contribution of interface atoms is negligible due to their full $sp^3$ character. Up to a strain of 0.40, the metallic character of the formed P1 polymorphs remains localized within the $sp^2$ sheets. However, as the phase transition progresses further, the metallic character of the resulting P2 phase is attributed to the initial interface atoms (indicated by the black arrow), where the $sp^2$ sheets are fully bonded to each other, causing some interface atoms to adopt an $sp^2$ configuration. Importantly, the diamond atoms remain strongly covalent bonding throughout the entire deformation process. These results demonstrate that the localized $sp^2$–$sp^3$ transformations within the $sp^2$ sheets lead to the formation of stable carbon polymorphs, which enable subsequent plastic response stage where the phase transition between carbon polymorphs continues to occur under large strains. This continuous phase transition effectively dissipates the strain energy and prevents the brittle fracture, thereby significantly enhancing fracture toughness.

**Conclusions**

In summary, we have successfully developed an extrinsic toughening strategy to synthesize a novel 3D-MWCNTs-diamond composite with a 3D continuous network heterogeneous structure under HPHT conditions. This composite exhibits an exceptional combination of high fracture toughness (36.4 MPa·m$^{1/2}$ as tested by SENB) and high hardness (91.6 ± 3.1 GPa as tested at 29.4 N). The remarkable performance is attributed to the robust interfacial bonding between MWCNTs and diamond, facilitated by mixed $sp^2$–$sp^3$ bonding, as well as the strategic incorporation of fibrous MWCNTs that bridge the gaps between diamond units to maintain the atomic interface continuity. The uneven stress distribution of diamond powder during high-pressure compression plays a pivotal role in achieving toughening through high-pressure interface engineering, allowing MWCNTs to bond with diamond while preserving their fibrous structure. Our findings, corroborated by advanced characterization techniques and DFT simulations, offer profound insights into the toughening mechanism, encompassing crack deflection, stress dispersion, and energy dissipation. This groundbreaking structural architecture strategy represents a significant milestone in the development of superhard materials and composite ceramics, promising cost reduction, lifetime extension, and future technical innovations across various practical applications.

**Experimental methods and tests.**

**Synthesis of 3D-MWCNTs-diamond bulks.** Commercially available high-purity diamond powder (grain size 0.8–1.3 μm, 99.99 % purity, Zhongnan Jiete Super abrasives Co., Zhengzhou, China) and multi-walled carbon nanotubes MWCNTs (inner diameter 5–10 nm, outer diameter 10–20 nm, length 10–20 μm, > 95% purity, Xianfeng Nanomaterials Technology Co., Nanjing, China) as starting materials (Supplementary Fig. 1a,b). The as-received MWNTs were dispersed either in the dispersant solution containing aromatic groups with the aid of ultrasonic agitation. The diamond powder was added to this solution and ultrasonically agitated, to result in either 95 wt% diamond + 5 wt% MWCNTs in the solution. The resulting slurries were ball-milled for 5 h (model 8000M, SPEX, USA), and were subsequently dried at 450 ºC in a resistance furnace (model KSL-1700X, Hefei Kejing, China) while being stirred. It was determined that the dispersant volatilizes at high temperature and does not cause any difference in the final composites in terms of their densities and microstructures.

The powder blends (Supplementary Fig. 1c) were packaged with Ta foil and pre-compressed into discs (3.5 mm in diameter and 2 mm in thickness) with a relative density of approximately 70%. HPHT experiments were performed with a 20-MN double-stage large-volume multi-anvil system with a standard assembly 14/8 (octahedron edge length /truncated edge length). Temperature was measured with type D W–Re thermocouples, and pressure was estimated from previously obtained calibration curves. The synthesized samples were ~3 mm in diameter and ~2 mm thick.

**Preparation of the tested 3D-MWCNTs-diamond samples.** The synthesized samples were first treated with acid to remove packaging and then polished at ambient pressure using a diamond paste. The polished sample will then be used to measure Vickers hardness and to process STEM specimens or micrometre-sized beams for the SENB tests. STEM specimens and the micrometre-sized beams were prepared using a focused ion beam (FIB, Helios G4 PFIB Cxe, Czekh). In the initial stage of the milling process, 1 nA current and 30 kV are used to dig two craters separated by thin walls. The wall was then taken out as a thin foil for STEM observation, or further processed into a test beam, with both ends connected to the bulk sample (Supplementary Fig. 6). A tungsten tip was used to extract the foil for STEM, and then, a lower current ranging from 0.1 nA to 10 pA to further polish the sample to the desired dimensions.

**Sample characterization**

The Diamond/MWNT powder blend was observed by scanning electron microscope (SEM) of FIB to investigate the dispersion of the MWNTs and mixing. The hardness of samples was characterized by a Vickers hardness tester (FV–700B, Future Tech, Tokyo, Japan) with an applied load of 29.4 N for 15 s. The Vickers hardness was determined as the Equation (1):

$$H_\mathrm{v} = 1854.4 \frac{F}{L^2} \qquad (1)$$

where $F$ (in N) is the applied load and $L$ (in μm) is the average diagonal of the Vickers indentation. the indentation fracture toughness was calculated as the Equation (2)

$$K_{\text{IC}} = 0.0166(\frac{E}{H_\text{V}})^{0.5} \times \frac{F}{C^{1.5}} \quad (2)$$

where $E$ is the Young's modulus of Diamond as 1050 GPa[44], and $C$ is the average length of the crack measured from the indent center. STEM measurements were performed by an aberration-corrected scanning transmission electron microscope equipped with double Cs correctors (CEOS) for the condenser lens and objective lens (ARM-200F, JEOL, Tokyo, Japan). Annular bright-field (ABF) images were acquired at acceptance angles of 11.5–23.0 mrad. SENB tests were conducted in an SEM equipped with an in situ mechanical test instrument (Hysitron PI-85). We prepared micrometre-sized test beams (about 10 μm in length, 0.4 μm in width and 1.3 μm in height), each with a precut notch in the centre point of the underside, to minimize size effect, as shown in Supplementary Fig. 6. The details for the SENB tests and fracture toughness calculations are described elsewhere[5], and information about the dynamic testing processes can be found in Supplementary Movie 1. Tensile fracture test was conducted on an FEI Tecnai F30 TEM system using an electrical holder from PicoFemto. The foil of MWCNTs-diamond composite was welded onto a copper FIB holder with a diameter of 3 mm. A tungsten tip was used to contact the left side of the foil; and the piezo manipulator to apply a tensile force by moving at a rate of about 0.01 nm·s$^{-1}$ (Supplementary Movie 2).

**Density functional theory simulations**

The density functional theory simulations were performed using the Vienna ab initio simulation package (VASP) code[45], and the projector augmented wave (PAW) method[46] with local density approximation (LDA) in the form of Ceperley-Alder[47,48] as the exchange-correlation functional. The energy cut-off of 520 eV, the energy convergence criterion of 10$^{-6}$ eV/cell, and the force convergence criterion of 10$^{-3}$ eV/Å were adopted. The Monkhorst-Pack k-mesh gamma-centered grids of 9 × 13 × 1 was used[49]. The stress-strain response was determined via the ADAIS code[50]. The schematic diagrams showing atomic models and electronic structures involved in this work were all created by VESTA[51] and SPaMD[52] packages.

**Acknowledgements**
The authors gratefully ac-knowledge the financial support from the National Key Research and Development Program of China (NO. 2021YFA1400300 to X. Yu, NO. 2024YFA1209800 to K Chen). This research was supported by the National Natural Science Foundation of China (NO. 51772011). The FIB-SEM characterization is supported by the Testing and Evaluation Center for High-performance Fibers at BUAA.

**Author contributions**
X.Y., J.Z., K.C., F.H. R.Z., K.Q., T.X, and X.S. conceived the project and designed the experiments. X.Y., J.Z., F.H., T.X., and X.S. carried out the design and fabrication of the 3D-MWCNTs. J.Z, K.Q., K.C., Z.W, and D.L. characterized all the samples. J.Z., K.Q., and K.C. performed the mechanical testing. K.C., K.Q., and J.Z. performed *in-situ* tensile tests. R.Z. and T.X. carried out MD simulation.

J.Z. and K.C. made the Movies. X.Y., K.C., J.Z., R.Z., F.H., L.W, D.H, H.G., J. L., F.L., and D.L. drafted the manuscript. All the authors discussed the results and commented on the manuscript.

**Corresponding author**
Email: yuxh@iphy.ac.cn; chenke0119@buaa.edu.cn; hongfang@iphy.ac.cn; zrf@buaa.edu.cn

**Competing interests**
The authors declare no competing financial interest.

**Additional information**
**Supplementary information** is available free of charge on the ……website at DOI:

**Data availability**
The authors declare that all data supporting the findings of this study are available within the paper and its Supplementary Information. Other supporting data are available from the corresponding author upon request.

**Author information**

**ORCID**
Jiawei Zhang: 0000-0002-0010-9933
Keliang Qiu: 0000-0002-5482-6671
Xi Shen: 0000-0003-4677-2455
Ke Chen: 0000-0001-7612-0226
Fang Hong: 0000-0003-0060-2063
Ruifeng Zhang: 0000-0002-9905-7271
Xiaohui Yu: 0000-0001-8880-2304

**References**

1. Yan, C. S. et al. Ultrahard diamond single crystals from chemical vapor deposition. *Phys. Status Solidi A* **201**, R25–R27 (2004).

2. Nie, A., Zhao, Z., Xu, B., & Tian, Y. Microstructure engineering in diamond-based materials. *Nat. Mater.* **949**, 1-14 (2025).

3. Novikov, N. V., Dub, S. N. & Malnev, V. I. Fracture toughness of diamond single crystals, *J. Hard Mater.* **2**, 3–11 (1991).

4. Dubrovinskaia, N., Dub, S. & Dubrovinsky, L. Superior wear resistance of aggregated diamond nanorods, *Nano Lett.* **6**, 824–826 (2006).

5. Yue, Y. et al. Hierarchically structured diamond composite with exceptional toughness. *Nature* **582**, 370–374 (2020).

6. Huang, Q. et al. Nanotwinned diamond with unprecedented hardness and stability. *Nature* **510**, 250–253 (2014).

7. Luo, K. et al. Coherent interfaces govern direct transformation from graphite to diamond. *Nature* **607**, 486–491 (2022).

8. Németh, P. et al. Diamond-graphene composite nanostructures. *Nano Lett.* **20**, 3611-3619 (2020).

9. Zeng, Z. et al. Synthesis of quenchable amorphous diamond. *Nat. Commun.* **8**, 322 (2017).

10. Lu, L., Chen, X., Huang, X. & Lu, K. Revealing the maximum strength in nanotwinned copper. *Science* **323**, 607–610 (2009).

11. Yue, Y., Zhang, Q., Yang, Z., Gong, Q. & Guo, L. Study of the mechanical behavior of radially grown fivefold twinned nanowires on the atomic scale. *Small* **12**, 3503–3509 (2016).

12. Lu, K. Stabilizing nanostructures in metals using grain and twin boundary architectures. *Nat. Rev. Mater.* **1**, 16019 (2016).

13. Yin, Z., Hannard, F. & Barthelat F. Impact-resistant nacre-like transparent materials. *Science* **364**, 1260–1263 (2019).

14. Chen, K. et al. Graphene oxide bulk material reinforced by heterophase platelets with multiscale interface crosslinking. *Nat. Mater.* **21**, 1121–1129 (2022).

15. Carré, H. & Daudeville, L. Load-bearing capacity of tempered structural glass. *J. Eng. Mech.* **125**, 914–921 (1999).

16. Li, F., Zhao, H., Yue, Y., Yang, Z., Zhang, Y. & Guo, L. Dual-phase super-strong and elastic ceramic. *ACS Nano*, **13**, 4191–4198 (2019).

17. Popov, V. N. Carbon nanotubes: properties and application. *Mater. Sci. and Eng.: R: Rep.* **43**, 61–102 (2004).

18. Jian, Q., Jiang, Z., Han, Y., Zhu, Y. & Li, Z. Fabrication and evaluation of mechanical properties of polycrystalline diamond reinforced with carbon-nanotubes by HPHT sintering. *Ceram. Int.* **46**, 21527–21532 (2020).

19. Yu, M. F., Yakobson, B. I. & Ruoff, R. S. Controlled sliding and pullout of nested shells in individual multiwalled carbon nanotubes. *J. Phys. Chem. B* **104**, 8764–8767 (2000).

20. Andrews, R. & Weisenberger, M. C. Carbon nanotube polymer composites. *Curr. Opin. Solid State Mater. Sci.* **8**, 31–37 (2004).

21. Green, M. J., Behabtu, N., Pasquali, M. & Adams, W. W. Nanotubes as polymers. *Polymer* **50**, 4979–4997 (2009).

22. Curtin, W. A. & Sheldon, B. W. CNT-reinforced ceramics and metals. *Mater. Today* **7**, 44–49 (2004).

23. Goh, P. S., Ismail, A. F. & Ng, B. C. Directional alignment of carbon nanotubes in polymer matrices: Contemporary approaches and future advances. *Compos. A Appl. Sci. Manuf.* **56**, 103–126 (2014).

24. Munch, E. et al. Tough, bio-inspired hybrid materials. *Science* **322**, 1516–1520 (2008).

25. Park, R. & Paulay, T. Reinforced concrete structures (John Wiley & Sons, 1975).

26. Zhang, J. et al. Transparent diamond ceramics from diamond powder. *J. Eur. Ceram. Soc.* **43**, 853–861(2022).

27. Guan, S. et al. Fragmentation and stress diversification in diamond powder under high pressure. *J. Appl.*

*Phys.* **124**, 215902 (2018).

28. Pan, Y. et al. Extreme mechanical anisotropy in diamond with preferentially oriented nanotwin bundles. *Proc. Natl. Acad. Sci.* **118**, e2108340118 (2021).
29. Li, Z. et al. Ultrastrong conductive in situ composite composed of nanodiamond incoherently embedded in disordered multilayer graphene. *Nat. Mater.* **22**, 42–49 (2023).
30. Shin, Y. A. et al. Nanotwin-governed toughening mechanism in hierarchically structured biological materials. *Nat. Commun.* **7**, 10772 (2016).
31. Ashby, M. F. Materials Selection in Mechanical Design 5th edn (Elsevier, 2017).
32. Tian, Y. et al. Ultrahard nanotwinned cubic boron nitride. *Nature* **493**, 385–388 (2013).
33. Tatarko, P. et al. Toughening effect of multi-walled boron nitride nanotubes and their influence on the sintering behaviour of 3Y-TZP zirconia ceramics. *J. Eur. Ceram. Soc.* **34**, 1829–1843 (2014).
34. Xu, F. M., Zhang, Z. J., Shi, X. L., Tan, Y. & Yang, J. M. Effects of adding yttrium nitrate on the mechanical properties of hot-pressed AlN ceramics. *J. Alloys Compd.* **509**, 8688–8691 (2011).
35. Liu, C. et al. Texture, microstructures, and mechanical properties of AlN-based ceramics with $Si_3N_4$–$Y_2O_3$ additives. *J. Am. Ceram. Soc.* **100**, 3380–3384 (2017).
36. Ritchie, R. O. The conflicts between strength and toughness. *Nat. Mater.* **10**, 817–822 (2011).
37. Gale, W. F. & Totemeier, T. C. Smithells Metals Reference Book 8th edn (Elsevier Butterworth-Heinemann, 2004).
38. Wang, H. et al. Nanostructured diamond-TiC composites with high fracture toughness. *J. Appl. Phys.* **113**, 043505 (2013).
39. Zhao, Y. *et al.* Enhancement of fracture toughness in nanostructured diamond–SiC composites. *Appl. Phys. Lett.* **84**, 1356–1358 (2004).
40. Zhao, B. et al. Enhanced strength of nano-polycrystalline diamond by introducing boron carbide interlayers at the grain boundaries. *Nanoscale Adv.* **2**, 691–698 (2020).
41. Li, B., Sun, H. & Chen, C. Large indentation strain-stiffening in nanotwinned cubic boron nitride. *Nat. Commun.* **5**, 4965 (2014).
42. Zhang, M. et al. Superhard $BC_3$ in cubic diamond structure. *Phys. Rev. Lett.* **114**, 015502 (2015).
43. Xu, T., Zheng, Z., Legut, D. & Zhang, R. Dinitrogen bonding induced metal-semiconductor transition leading to ultrastiffening in boron subnitride. *Phys. Rev. B* **108**, L180103 (2023).
44. Weidner, D. J., Wang, Y. & Vaughan, M. T. Strength of diamond. *Science* **266**, 419–422 (1994).
45. Kresse, G. & Furthmüller, J. Efficient iterative schemes for ab initio total-energy calculations using a plane-wave basis set. *Phys. Rev. B* **54**, 11169–11186 (1996).
46. Kresse, G. & Joubert, D. From ultrasoft pseudopotentials to the projector augmented-wave method. *Phys. Rev. B* **59**, 1758–1775 (1999).
47. Ceperley, D. M. & Alder, B. J. Ground state of the electron gas by a stochastic method. *Phys. Rev. Lett.* **45**, 566–569 (1980).
48. Perdew, J. P. & Zunger, A. Self-interaction correction to density-functional approximations for many-body systems. *Phys. Rev. B* **23**, 5048–5079 (1981).
49. Monkhorst, H. J. & Pack, J. D. Special points for Brillouin-zone integrations. *Phys. Rev. B* **13**, 5188–5192 (1976).
50. Zhang, S. H., Fu, Z. H. & Zhang, R. F. ADAIS: Automatic derivation of anisotropic ideal strength via

high-throughput first-principles computations. *Comput. Phys. Commun.* **238**, 244–253 (2019).
51. Momma, K. & Izumi, F. VESTA 3 for three-dimensional visualization of crystal, volumetric and morphology data. *J. Appl. Cryst.* **44**, 1272–1276 (2011).
52. Liu, Z. R., Yao, B. N. & Zhang, R. F. SPaMD studio: An integrated platform for atomistic modeling, simulation, analysis, and visualization. *Comp. Mater. Sci.* **210**, 111027 (2022).

# Supplementary Information

## Three-Dimensional Continuous Multi-Walled Carbon Nanotubes Network-Toughened Diamond Composite

Jiawei Zhang[1,2], Keliang Qiu[3], Tengfei Xu[4], Xi Shen[1], Junkai Li[5], Fengjiao Li[1], Richeng Yu[1], Huiyang Gou[5], Duanwei He[6], Liping Wang[7], Zhongzhou Wang[3], Guodong Li[8], Yusheng Zhao[2], Ke Chen[3,8,*], Fang Hong[1,*], Ruifeng Zhang[4,*], Xiaohui Yu[1,*]

1. Beijing National Laboratory for Condensed Matter Physics, Institute of Physics, Chinese Academy of Sciences, Beijing 100190, P. R. China.
2. Eastern Institute for Advanced Study, Eastern Institute of Technology; Ningbo 315201, P. R. China.
3. School of Chemistry, Beihang University; Beijing 100191, P. R. China.
4. School of Materials Science and Engineering, Beihang University; Beijing 100191, P. R. China.
5. Center for High Pressure Science and Technology Advanced Research; Beijing 100193, P. R. China.
6. Institute of Atomic and Molecular Physics, Sichuan University; Chengdu 610065, P. R. China.
7. Academy for Advanced Interdisciplinary Studies, and Department of Physics, Southern University of Science and Technology; Shenzhen 518055, P. R. China.
8. State Key Lab of Tropic Ocean Engineering Materials and Materials Evaluation, Hainan University, Haikou 570228, P. R. China.
9. These authors contributed equally: Jiawei Zhang, Keliang Qiu, Tengfei Xu, Xi Shen,

**Email**: yuxh@iphy.ac.cn; chenke0119@buaa.edu.cn; hongfang@iphy.ac.cn; zrf@buaa.edu.cn

The file includes:

Supplementary Text
Figs. S1 to S12
Tables S1

Other Supplementary Materials for this manuscript include the following:

Movies S1 and S2

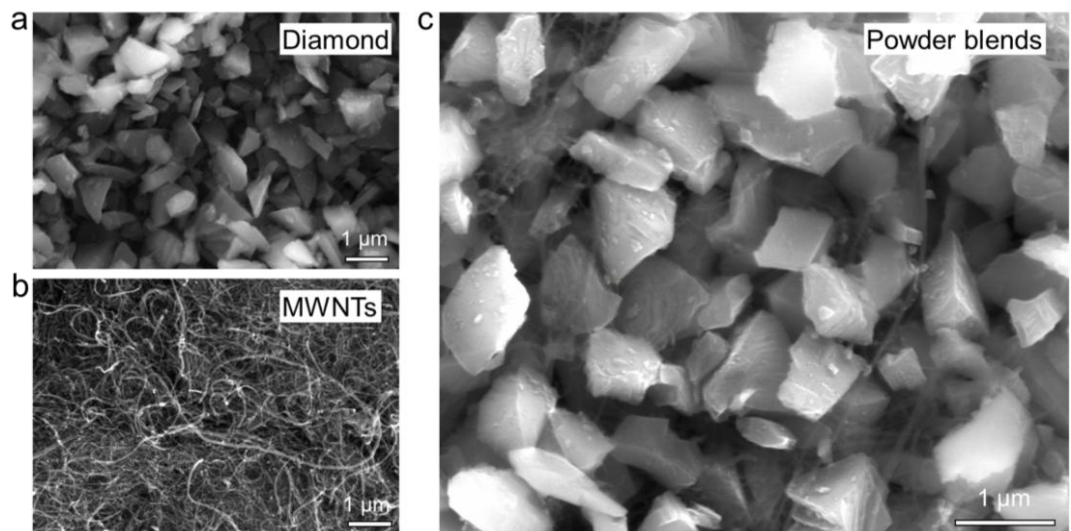

**Supplementary Fig. 1 | Characterization of the starting materials**. **a**, SEM image of the diamond powder with grain size of 0.8–1.3 μm. **b**, SEM image of MWCNTs with large aspect ratio. **c**, SEM image of diamond powders blended with MWCNTs.

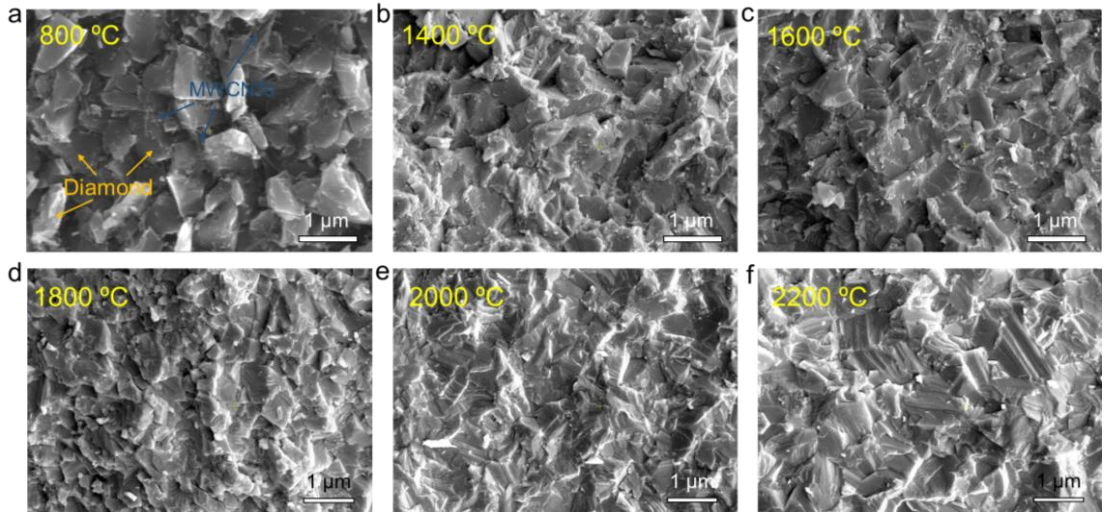

**Supplementary Fig. 2 | Characterization for fractured surfaces of 3D-MWCNTs-diamond composites sintered at different temperatures (from 800 °C to 2200 °C) under 15 GPa**. **a-d**, SEM images for the fracture morphologies of the composites sintered at 800 °C (a), 1400 °C (b), 1600 °C (c), and 1800 °C (d), respectively. The composites are gradually densified as the sintering temperature increases from 1400 °C to 1800 °C, but there are still a large number of pores, in which MWCNTs are protected from the pressure compression. Fine diamond grains are produced by the fracture of large diamond grains, which is also the main densification behavior at lower sintering temperature. **e,f**, SEM images for the fracture morphologies of the composites sintered at 2000 °C (e) and 2200 °C (f). The diamond grains undergo obvious plastic deformation and achieve good bonding when the temperature is equal to (higher than) 2000 °C. The pores are greatly reduced or even eliminated, and plastic deformation becomes the main densification behavior at higher sintering temperature (≥2000 °C).

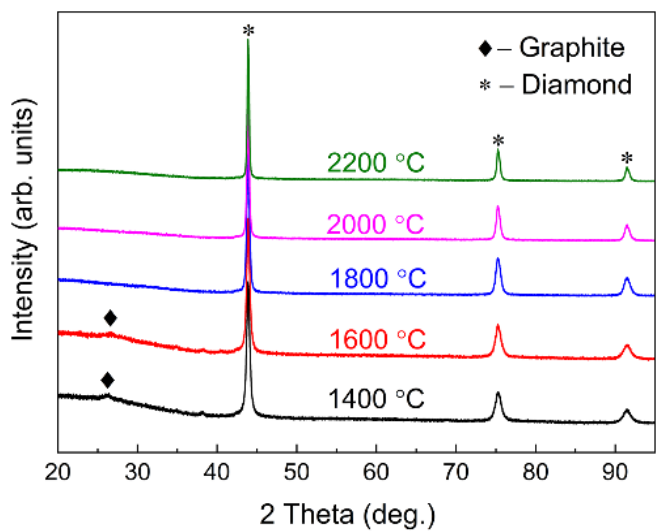

**Supplementary Fig. 3 | XRD patterns of 3D-MWCNTs-diamond composites sintered at different temperatures under 15 GPa.** The graphite peak gradually weakens and disappears as the sintering temperature increases from 1400 °C to 2200 °C. MWCNTs are gradually transformed into diamond with smaller pores and greater actual pressure due to fracture and plastic deformation of diamond particles. A small amount of graphite phase may also exist at 1800 °C and 2000 °C, but it cannot be detected by the low intensity X-ray source.

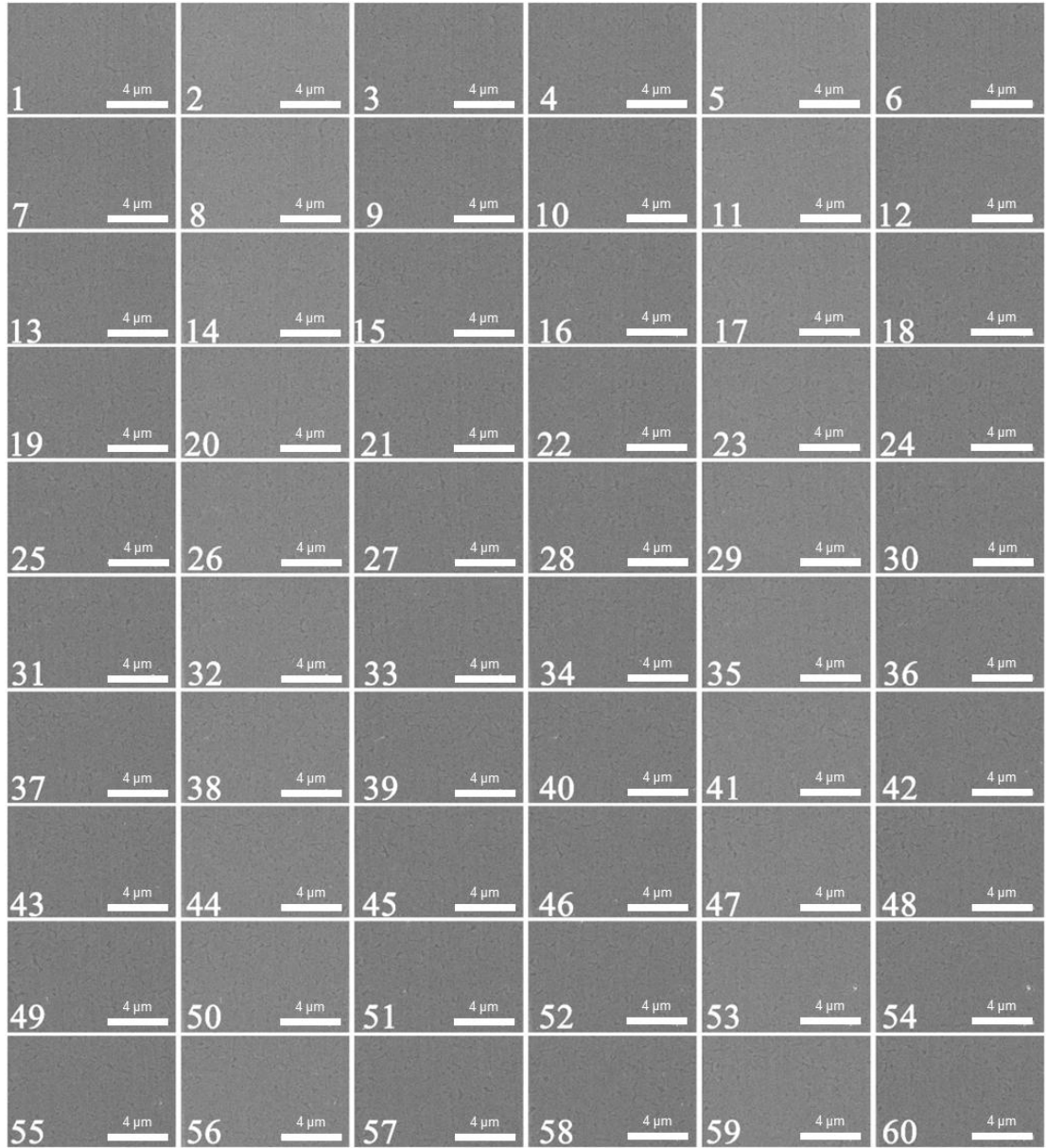

**Supplementary Fig. 4 | The SEM image of continuous sixty composite slices (size: 10 μm × 6 μm × 20 nm), obtained by focused ion beam (FIB) technology**. The SEM images show uniform defect network structure (dark regions) around diamond grains on the 2D planes, preliminarily demonstrating the formation of 3D MWCNTs network structure around diamond grains.

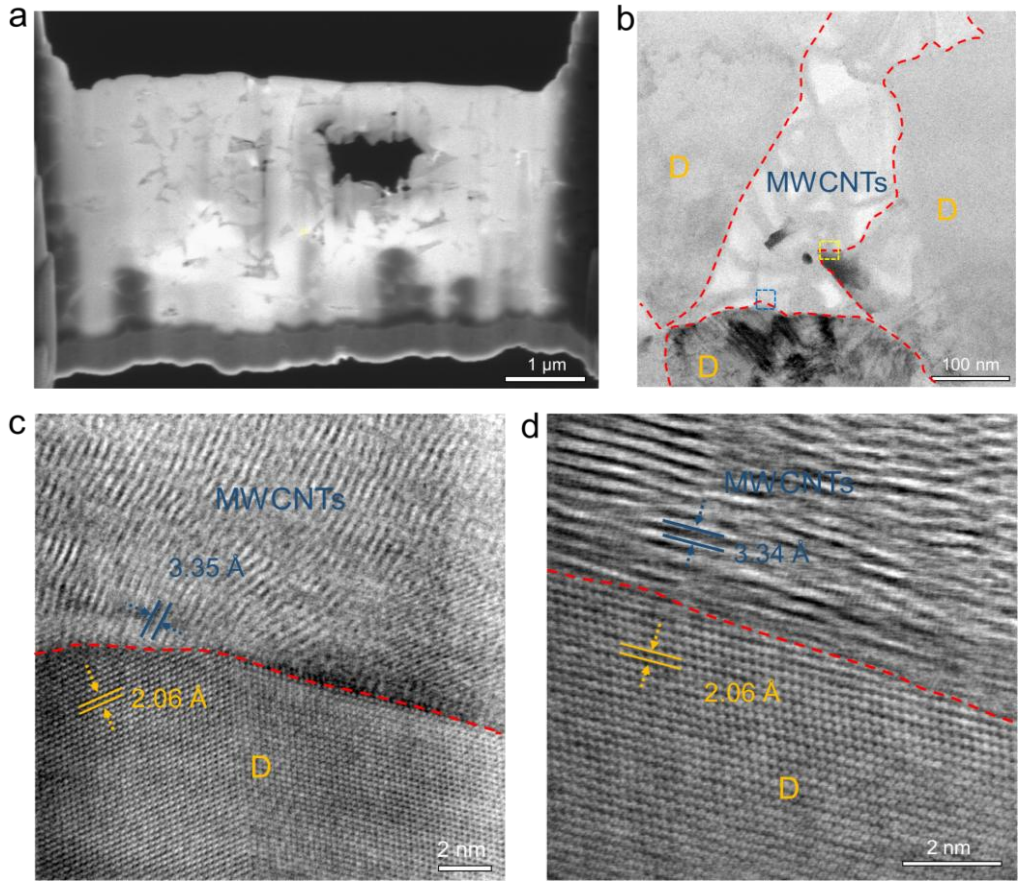

**Supplementary Fig. 5 | Microstructure characterization of 3D-MWCNTs-diamond composite prepared at 15 GPa and 2000 °C, revealed by STEM. a**, A low-magnified STEM image of a thin foil taken out to prepare STEM sample. The foil is translucent and the gap of diamond grains can be vaguely observed. **b**, A low-magnified ABF-STEM image of the thin foil, showing that MWCNTs are filled in the gaps between diamond grains, forming a 3D continuous MWCNTs network in the polycrystal diamond matrix. **c,d**, High-magnified ABF-STEM images of heterogenous MWCNTs/diamond interfaces, corresponding to the yellow box (c) and blue box (d) in (b), respectively.

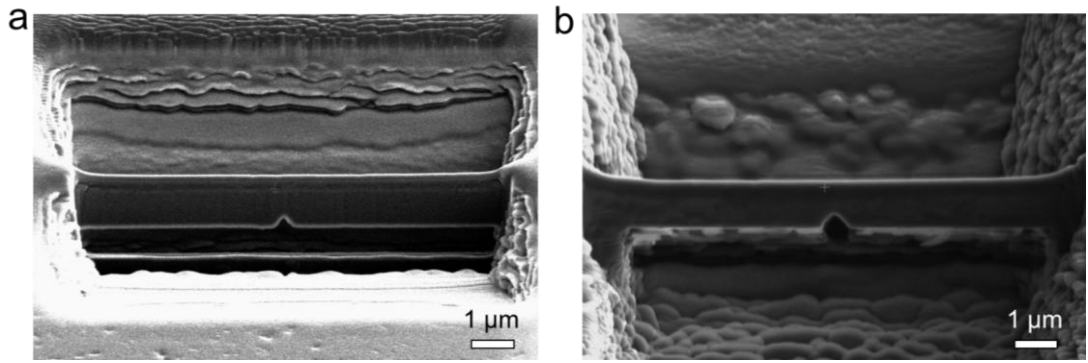

**Supplementary Fig. 6 | SEM images of the micrometre-sized beams with different pre-existing cracks in 3D-MWCNTs-diamond for SENB testing.**

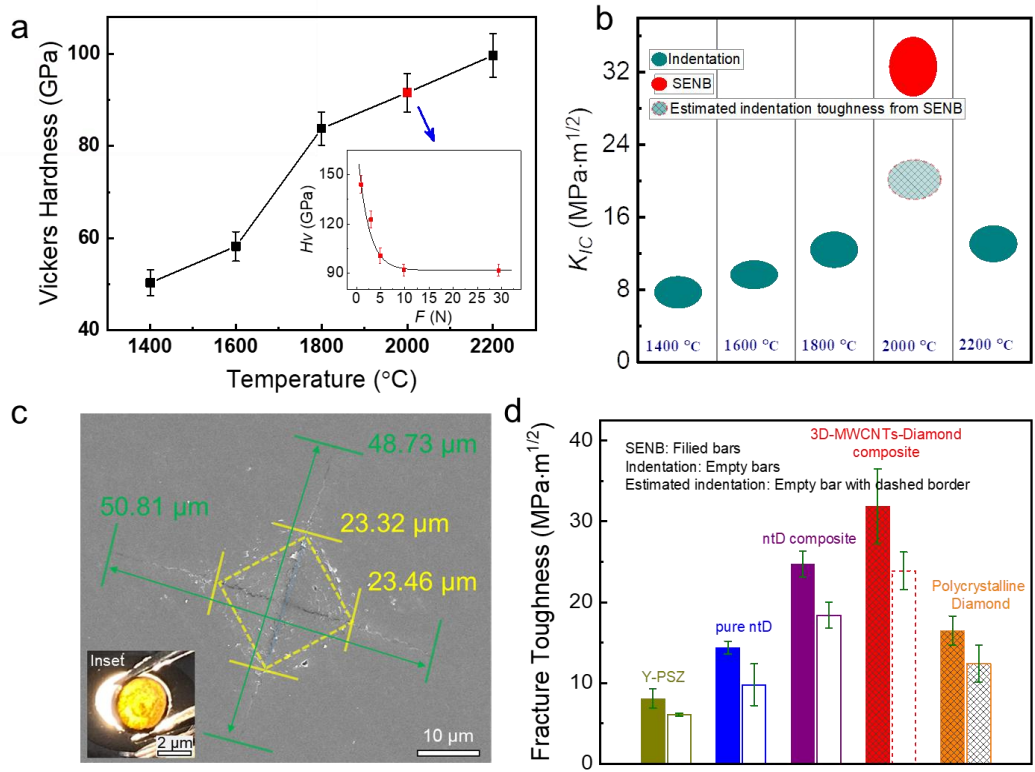

**Supplementary Fig. 7 | Vickers hardness and fracture toughness of 3D-MWCNTs-diamond composites, in comparison with other materials**. **a**, Vickers hardness of the composites sintered at different temperatures (1400 °C~2200 °C) under 15 GPa. Inset in **a**: Vickers hardness of the composites (15 GPa, 2000 °C) as a function of the applied load. Beyond 9.8 N, Vickers hardness decreases to the asymptotic values of ~90 GPa. **b**, Fracture toughness of the composites sintered at different temperatures under 15 GPa. The fracture toughness in **b** is cyan from the indentation test and red from the SENB test. **c**, The typical post-indentation SEM micrograph of the composite surface prepared at 15 GPa and 2200 °C, subjected to a 29.4 N load by a Vicker indentation test, showing visible indentation cracks. Inset: photograph of the composite, transparent under the bottom light source. **d**, Comparison of fracture toughness values from SENB and indentation measurements for commercial yttria partially stabilized zirconia (Y-PSZ, Beilong Electronics, China), pure nanotwinning diamond (pure ntD), ntD composite, polycrystalline diamond, and 3D-MWCNTs-diamond composite. SENB, filled bars; indentation: empty bars; Estimated indentation: empty bars with dashed border. The loads for indentation fracture toughness measurement were 49 N for Y-PSZ, and 19.6 N for pure nt-diamond and nt-diamond composite (References). Error bars indicate 1 s.d. ($n = 5$ for Y-PSZ and nt-diamond composite, $n = 3$ and 5 for indentation and SENB measurements of pure nt-diamond, respectively). ntD, nt-diamond, polycrystalline diamond, and 3D-MWCNTs-diamond composite.

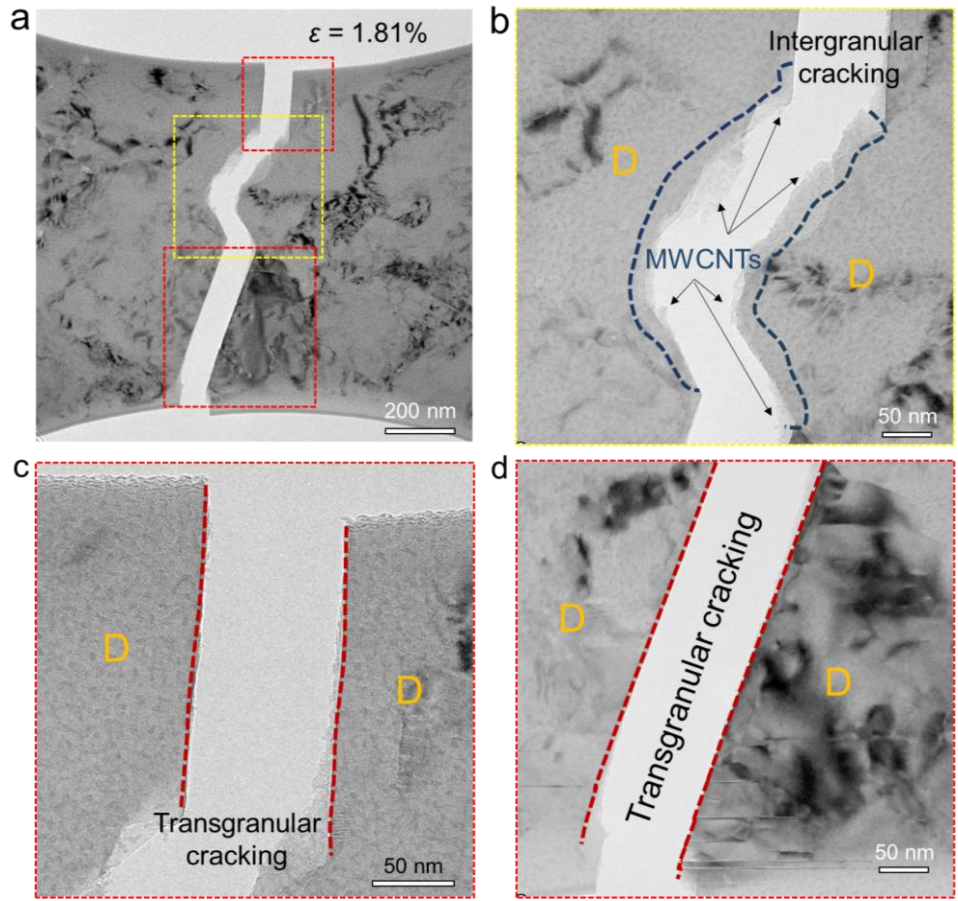

**Supplementary Fig. 8 | The fracture behavior of a specially designed (bone-like) 3D-MWCNTs-diamond composite sample, by an *in situ* tensile test in the TEM. a,** TEM image of fractured 3D-MWCNTs-diamond composite slice, corresponding to the TEM image of Fig. 4a ($\varepsilon$ = 1.81%) in the main manuscript. **b,** Enlarged TEM image of fractured 3D-MWCNTs-diamond composite in intergranular cracking region, taken from the yellow dashed frame in a. **c,d,** Enlarged TEM images of fractured 3D-MWCNTs-diamond composite in transgranular cracking region, taken from the red dashed frames in a.

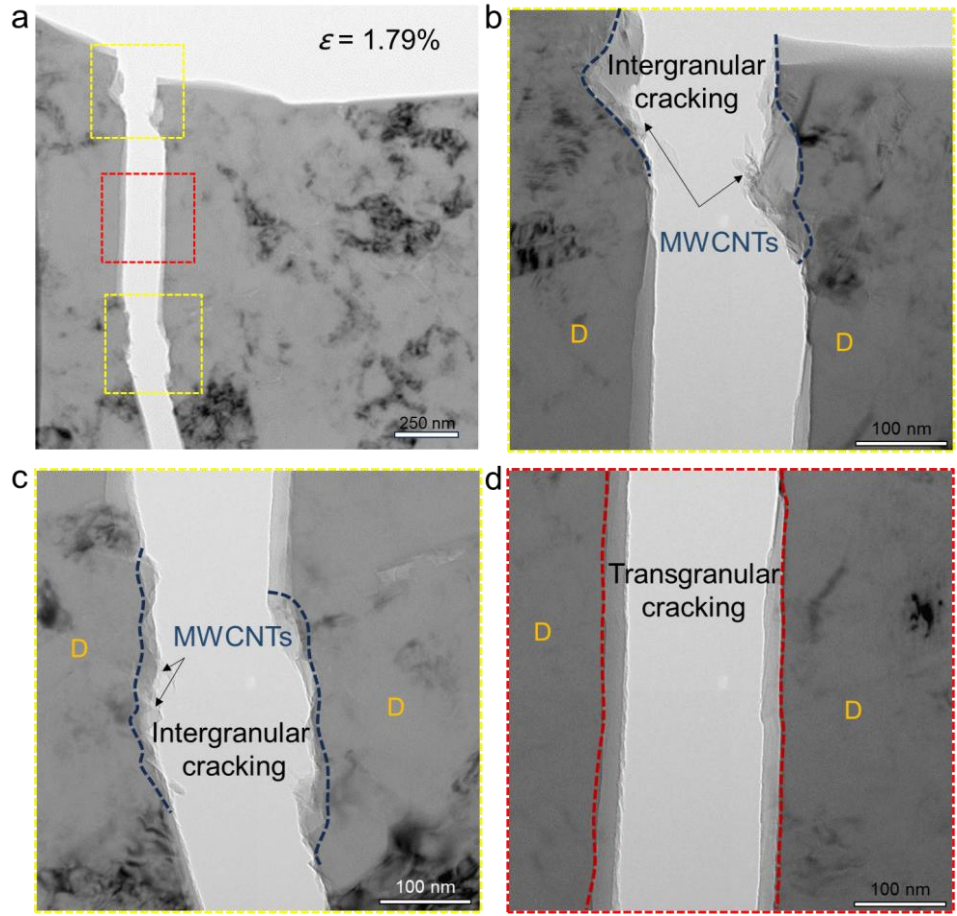

**Supplementary Fig. 9 | The fracture behavior of a specially designed (I-shaped) 3D-MWCNTs-diamond composite sample, by an *in situ* tensile test in the TEM. a,** TEM image of fractured 3D-MWCNTs-diamond composite slice, at the strain of *ε* = 1.79%. **b,c,** Enlarged TEM images of fractured 3D-MWCNTs-diamond composite in intergranular cracking region, taken from the yellow dashed frames in a. **d**, Enlarged TEM image of fractured 3D-MWCNTs-diamond composite in transgranular cracking region, taken from the red dashed frame in a.

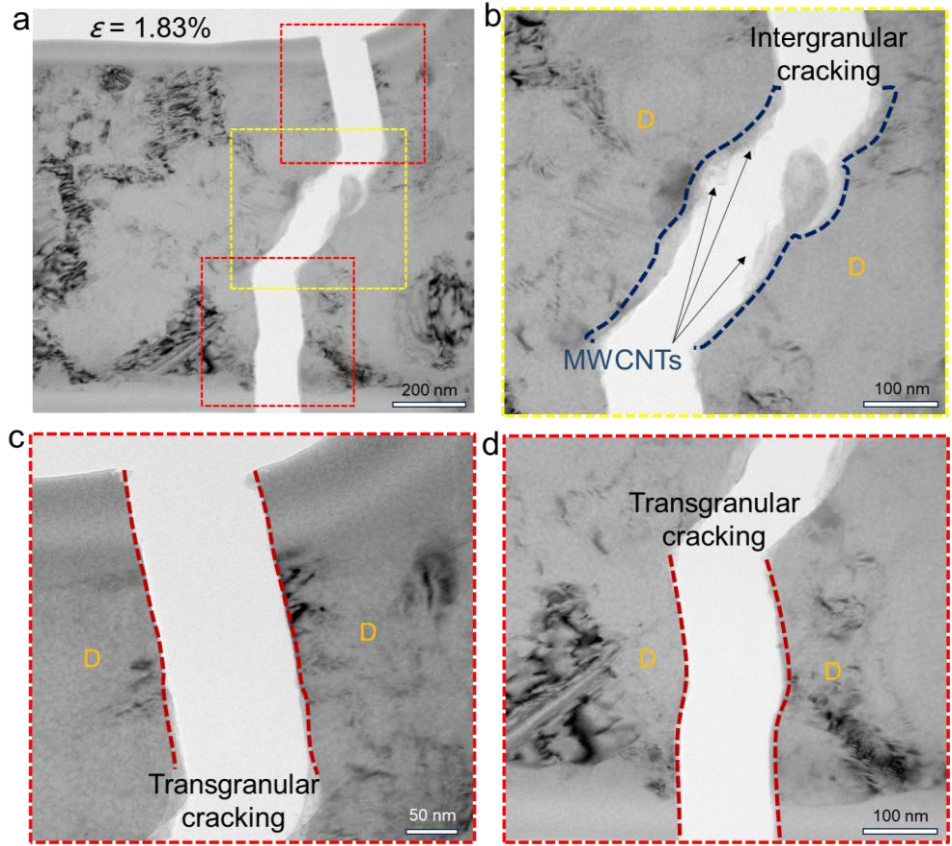

**Supplementary Fig. 10 | The fracture behavior of a specially designed (I-shaped) 3D-MWCNTs-diamond composite sample, by an *in situ* tensile test in the TEM. a,** TEM image of fractured 3D-MWCNTs-diamond composite slice, at the strain of $\varepsilon = 1.83\%$. **b,** Enlarged TEM image of fractured 3D-MWCNTs-diamond composite in intergranular cracking region, taken from the yellow dashed frame in a. **c,d**, Enlarged TEM images of fractured 3D-MWCNTs-diamond composite in transgranular cracking region, taken from the red dashed frames in a.

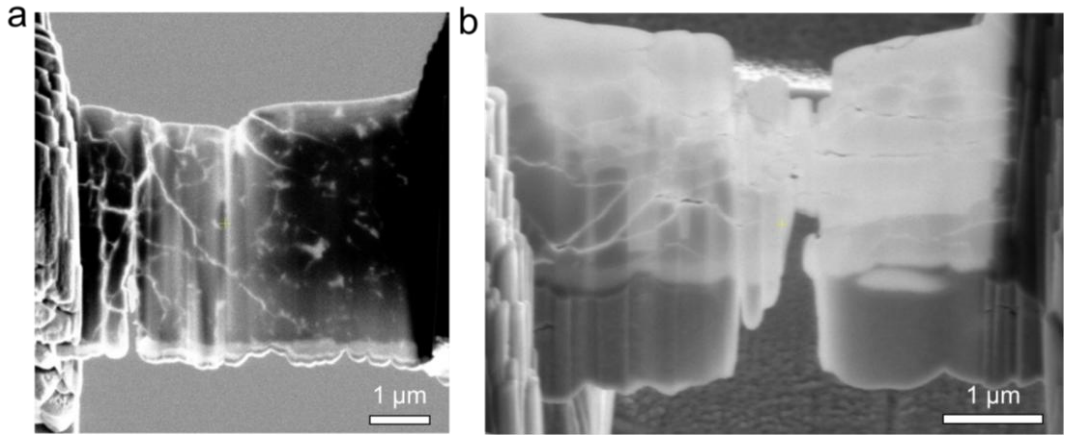

**Supplementary Fig. 11 | Morphology characterization of 3D-MWCNTs-diamond samples for micromechanical tests, obtained by a focused ion beam (FIB) technology. a,b** SEM images of thin composite slices with cracks under different forces, measured by Vickers indentation.

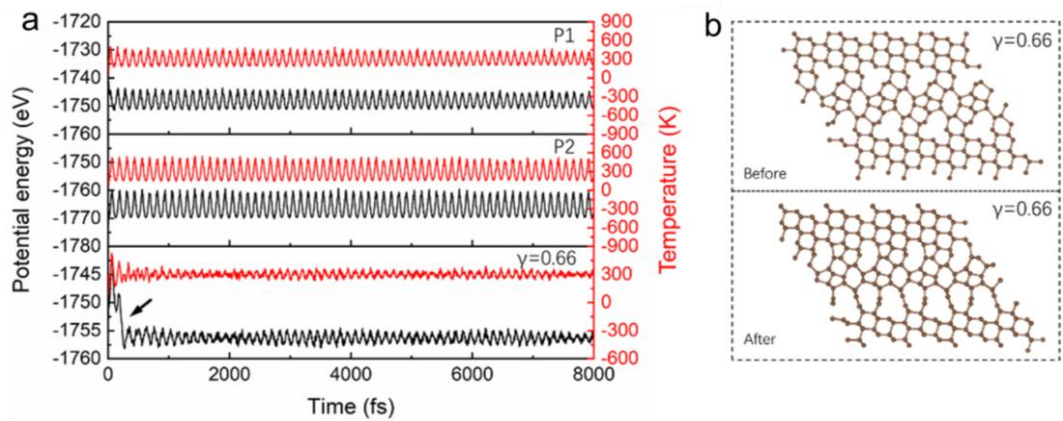

**Supplementary Fig. 12 | The Ab initio molecular dynamics (AIMD) simulation results of several carbon polymorphs**. **a**, The variation of total potential energy and temperature fluctuation for the supercells of different polymorphs during AIMD simulation of the Canonical ensemble (NVT) at 300 K. **b**, The deformed structures at the strain of 0.66 before and after the AIMD simulation. Different from the structural stability of P1 and P2 phases, for structure at strain of 0.66, there is a noticeable drop in energy (see black arrow) and a significant structural change before and after the relaxation, suggesting the instability of this phase-transitioned structure and thus the final failure of the composite during the deformation.

**Supplementary Table 1**. Fracture toughness of 3D-MWCNTs-diamond composite samples (synthesized at 15 GPa and 2000 °C) from SENB tests.

| Sample No. | Fracture toughness (MPa•m$^{1/2}$) | | |
|---|---|---|---|
| | measured value | average value | median value |
| 1 | 27.3 | | |
| 2 | 26.7 | | |
| 3 | 36.4 | 31.9 ± 4.6 | 33.8 |
| 4 | 35.5 | | |
| 5 | 33.8 | | |

**Supplementary Movie 1**.

Multi-cycle bending test on 3D-MWCNTs-diamond (shown in Fig. 2b-f). Video speed at 8 times the speed of experiment.

**Supplementary Movie 2**.

*in-situ* tensile fracture test on 3D-MWCNTs-diamond. Video speed at 100 times the speed of experiment.